\documentclass[fleqn,10pt]{wlscirep}
\usepackage[utf8]{inputenc}
\usepackage[T1]{fontenc}

\title{Lightning Network: a second path towards centralisation of the Bitcoin economy}

\author[1,2]{Jian-Hong Lin}
\author[3]{Kevin Primicerio}
\author[4]{Tiziano Squartini}
\author[5]{Christian Decker}
\author[1,6]{Claudio J. Tessone}

\affil[1]{URPP Social Networks, University of Zurich, Andreasstrasse 15, CH-8050 Z\"urich (Switzerland)}
\affil[2]{ETH Z\"urich, Department of Management, Technology and Economics, Scheuchzerstrasse 7, CH-8092 Z\"urich (Switzerland)}
\affil[3]{Mathmatiques et Informatique pour la Complexité et les Systèmes, CentraleSupélec, Université Paris-Saclay, FR-91190 Gif-Sur-Yvette (France)}
\affil[4]{IMT School for Advanced Studies Lucca, Piazza San Francesco 19, I-55100 Lucca (Italy)\thanks{tiziano.squartini@imtlucca.it}}
\affil[5]{Blockstream Corporation Inc., QC H4M 2X6 Montreal (Canada)}
\affil[6]{UZH Blockchain Center, University of Zurich, Andreasstrasse 15, CH-8050 Z\"urich (Switzerland)}

\begin{abstract}
The Bitcoin Lightning Network (BLN), a so-called ``second layer'' payment protocol, was launched in 2018 to scale up the number of transactions between Bitcoin owners. In this paper, we analyse the structure of the BLN over a period of 18 months, ranging from $12^{th}$ January 2018 to $17^{th}$ July 2019, at the end of which the network has reached 8.216 users, 122.517 active channels and 2.732,5 transacted bitcoins. Here, we consider three representations of the BLN: the \emph{daily snapshot} one, the \emph{weekly snapshot} one and the \emph{daily-block snapshot} one. By studying the topological properties of the binary and weighted versions of the three representations above, we find that the total volume of transacted bitcoins approximately grows as the square of the network size; however, despite the huge activity characterising the BLN, the bitcoins distribution is very unequal: the average Gini coefficient of the node strengths (computed across the entire history of the Bitcoin Lightning Network) is, in fact, $\simeq 0.88$ causing the $10\%$ ($50\%$) of the nodes to hold the $80\%$ ($99\%$) of the bitcoins at stake in the BLN (on average, across the entire period). This concentration brings up the question of which minimalist network model allows us to explain the network topological structure. Like for other economic systems, we hypothesise that local properties of nodes, like the degree, ultimately determine part of its characteristics. Therefore, we have tested the goodness of the Undirected Binary Configuration Model (UBCM) in reproducing the structural features of the BLN: the UBCM recovers the disassortative and the hierarchical character of the BLN but underestimates the centrality of nodes; this suggests that the BLN is becoming an increasingly centralised network, more and more compatible with a core-periphery structure. Further inspection of the resilience of the BLN shows that removing hubs leads to the collapse of the network into many components, an evidence suggesting that this network may be a target for the so-called \emph{split attacks}.
\end{abstract}

\begin{document}

\flushbottom
\maketitle

\thispagestyle{empty}

\section*{Introduction}

The gain of popularity of Bitcoin \cite{nakamoto2019bitcoin} has manifested the problems related to the scalability of the technology upon which it is based: in fact, only a limited amount of transactions per second - whose number is proportional to the size of a block and its release frequency - can be processed by Bitcoin. This shortcoming may  prevent the adoption of this payment network at a global scale, especially when considering that classic payment mechanisms (e.g. traditional credit cards) are able to achieve tens of thousands of transactions per second. A na\"ive (and short term) solution would be represented by an increase of the block size: larger blocks, however, would require larger validation time, storage capability and bandwidth costs, in turn favouring \emph{centralisation}, as fewer entities would become able to validate the new blocks that are appended to the Blockchain; moreover, centralisation in the validation process would make the system less resilient, i.e. more prone to faults and attacks.

The Bitcoin Lightning Network (BLN) \cite{poon2016bitcoin,bartolucci2019percolation,seres2019topological} aims at breaking the trade-off between block size and centralisation by processing most of the transactions off-chain: it is a ``Layer 2'' protocol that can operate on top of Blockchain-based cryptocurrencies such as Bitcoin. The origin of the BLN can be traced back to the birth of Bitcoin itself, as an attempt to create \emph{payment channels} across which any two users could exchange money without burdening the entire network with their transaction data - thus allowing for \emph{cheaper} and \emph{faster} transactions (as both the mining fees and the Blockchain confirmation are no longer required). The BLN has, thus, promised to represent a solution to the Bitcoin scalability problem that does not sacrifice the key feature of Bitcoin itself, i.e. \emph{decentralisation}, that characterises its \emph{architecture} (i.e. the number of computers constituting the network), its \emph{political organisation} (i.e. the number of individuals controlling the network) and its \emph{wealth distribution} (i.e. the number of individuals owning the actual supply), while enhancing the circulation and the exchange of the native assets.

What remains unclear, however, is if the promise has been fulfilled. Several contributions have pointed out that the BLN structure can be ameliorated to improve its security, by testing its robustness against the so-called topology-based attacks \cite{seres2019topological,rohrer2019discharged} and comparing them with the effects of random failures. What emerges from these analyses is that the BLN can indeed be disrupted upon carefully choosing the nodes to isolate - thus compromising the nodes reachability, the payment success ratio, etc. As these findings can be related to the way the BLN topological structure has evolved, this paper is devoted to answer a simple question: \emph{has the Bitcoin Lightning Network become increasingly centralized?} To  provide an answer, we consider the BLN payment channels across a period of 18 months, i.e. from $12^{th}$ January 2018 to $17^{th}$ July 2019, and analyze it at both the daily and the weekly timescale.

What emerges is that the BLN is characterised by an unequal wealth distribution and by a larger-than-expected centrality of nodes, thus suggesting that the BLN indeed suffers from the aforementioned centralisation issue.

\section*{Methods}

\paragraph{Notation.}For each time snapshot $t$, the BLN can be described as a weighted, undirected network with total number of nodes $N^{(t)}$ and represented by the $N^{(t)}\times N^{(t)}$ symmetric matrix $\mathbf{W}^{(t)}$~\cite{newman2003structure,newman2006structure} whose generic entry $w_{ij}^{(t)}$ indicates the total amount of money exchanged between $i$ and $j$, across all channels, at time $t$. The total amount of money exchanged by node $i$, at time $t$, is $s_{i}^{(t)}=\sum_{j(\neq i)=1}^{N^{(t)}}{w_{ij}^{(t)}}$, a quantity that will be also called \emph{capacity}. For the present analysis, we also consider the BLN binary adjacency matrix $\mathbf{A}^{(t)}$, whose generic entry reads $a_{ij}^{(t)}=1$ if $w_{ij}^{(t)}>0$ and $a_{ij}^{(t)}=0$ otherwise. Naturally, the presence of a link between any two nodes $i$ and $j$, i.e. $a_{ij}^{(t)}=1$, indicates that one or more payment channels are open, between the same nodes, at time $t$ and the total number of open channels (i.e. links) is simply provided by $L^{(t)}=\sum_{i=1}^{N^{(t)}}\sum_{j=i+1}^{N^{(t)}}a_{ij}^{(t)}$.

\paragraph{Centrality measures.} Indices measuring the centrality of a node aim at quantifying the importance of a node in a network, according to some, specific topological property~\cite{bonacich1987power,borgatti2005centrality,newman2018networks,rodrigues2019network}. Among the measures proposed so far, of particular relevance are the \emph{degree centrality}, the \emph{closeness centrality}, the \emph{betweenness centrality} and the \emph{eigenvector centrality}. Let us briefly describe them:

\begin{itemize}
\item the degree centrality~\cite{newman2018networks,rodrigues2019network} of node $i$ coincides with the degree of node $i$, i.e. the number of its neighbours, normalized by the maximum attainable value, i.e. $N-1$:

\begin{equation}\label{equation1}
k^c_{i}=\frac{k_i}{N-1}
\end{equation}
where $k_i=\sum_{j(\neq i)=1}^Na_{ij}$. From the definition above, it follows that the most central node, according to the degree variant, is the one connected to all the other nodes;

\item the closeness centrality~\cite{newman2018networks,rodrigues2019network} of node $i$ is defined as

\begin{equation}\label{equation3}
c^c_{i}=\frac{N-1}{\sum_{j(\neq i)=1}^Nd_{ij}}
\end{equation}
where $d_{ij}$ is the topological distance between nodes $i$ and $j$, i.e. the length of the shortest path(s) connecting them: in a sense, the closeness centrality answers the question ``how reachable is a given node?'' by measuring the length of the patterns that connect it to the other vertices. From the definition above, it follows that the most central node, according to the closeness variant, is the one lying at distance 1 by each other node;

\item the betweenness centrality~\cite{newman2018networks,newman2005measure,pfitzner2013betweenness,bonacich2007some} of node $i$ is given by

\begin{equation}\label{equation2}
b^c_{i}=\sum_{s(\neq i)=1}^N\sum_{t(\neq i,s)=1}^N\frac{\sigma_{st}(i)}{\sigma_{st}}
\end{equation}
where $\sigma_{st}$ is the total number of shortest paths between node $s$ and $t$ and $\sigma_{st}(i)$ is the number of shortest paths between nodes $s$ and $t$ that pass through node $i$. From the definition above, it follows that the most central node, according to the betweenness variant, is the one lying ``between'' any two other nodes;

\item the eigenvector centrality~\cite{newman2018networks,liu2016locating,bonacich2007some} of node $i$, $e^c_{i}$, is defined as the $i$-th element of the eigenvector corresponding to the largest eigenvalue of the binary adjacency matrix (whose existence is ensured by the Perron-Frobenius theorem). According to the definition above, a node with large eigenvector centrality is connected to other ``well connected'' nodes. In this sense, its behavior is similar to the PageRank centrality index.
\end{itemize}

\paragraph{Gini coefficient.} The Gini coefficient has been introduced to quantify the inequality of a country income distribution \cite{morgan1962anatomy,crucitti2006centrality}: it ranges between 0 and 1, with a larger Gini coefficient indicating a larger ``unevenness'' of the income distribution. Here, we apply it to both the distribution of the centrality measures of nodes, i.e.

\begin{equation}
G_c=\frac{\sum_{i=1}^N\sum_{j=1}^N|c_i-c_j|}{2N\sum_{i=1}^N{c_i}}
\end{equation}
where $c_i=k^c_i,c^c_i,b^c_i,e^c_i$ and to the distribution of the total amount of money exchanged by the nodes of the BLN, i.e.

\begin{equation}\label{equation6}
G_s=\frac{\sum_{i=1}^N\sum_{j=1}^N|s_i-s_j|}{2N\sum_{i=1}^N{s_i}}.
\end{equation}

\paragraph{Centralisation measures.} The centrality indices defined above are all normalized between 0 and 1 and provide a rank of the nodes of a network, according to the topological feature chosen for their definition. Sometimes, however, it is useful to compactly describe a certain network structure in its entirety. To this aim, a family of indices has been defined (the so-called \emph{centralisation indices}), encoding the comparison between the structure of a given network and that of the reference network, according to the chosen index. In mathematical terms, any centralisation index reads $C_c=\frac{\sum_{i=1}^N(c^*-c_i)}{\max\{\sum_{i=1}^N(c^*-c_i)\}}$, where $c^*=\max\{c_i\}_{i=1}^N$ represents the maximum value of the chosen centrality measure computed over the network under consideration and the denominator is calculated over the benchmark, defined as the graph providing the maximum attainable value of the quantity $\sum_{i=1}^N(c^*-c_i)$. As it can be proven that the most centralized structure, according to the degree, closeness and betweenness centrality, is the \emph{star graph}, one can define the corresponding centralisation indices:

\begin{itemize}
\item the \emph{degree-centralisation} index, as
\begin{equation}
C_{k^c}=\frac{\sum_{i=1}^N(k^*-k^c_i)}{(N-1)(N-2)};
\end{equation}
\item the \emph{closeness-centralisation} index, as 
\begin{equation}
C_{c^c}=\frac{\sum_{i=1}^N(c^*-c^c_i)}{(N-1)(N-2)/(2N-3)};
\end{equation}
\item the \emph{betweenness-centralisation} index, as
\begin{equation}
C_{b^c}=\frac{\sum_{i=1}^N(b^*-b^c_i)}{(N-1)^2(N-2)/2};
\end{equation}

\item the \emph{eigenvector-centralisation} index, as

\begin{equation}
C_{e^c}=\frac{\sum_{i=1}^N(e^*-e^c_i)}{(\sqrt{N-1}-1)(N-1)/(\sqrt{N-1}+N-1)}.
\end{equation}
\end{itemize}

For what concerns the eigenvector index, the star graph does not represent the maximally centralised structure: however, we keep it for the sake of homogeneity with the other quantities.

\paragraph{Benchmarking the observations.} Beside providing an empirical analysis of the BLN, in what follows we will also benchmark our observations against a model discounting available information to some extent. Like for other economic and financial systems, we hypothesise that local properties of nodes ultimately determine the BLN structure: specifically, we focus on the \emph{degrees} and adopt the the Undirected Binary Configuration Model (UBCM) as a reference model \cite{park2004statistical,squartini2011analytical}. The UBCM  captures the idea that the probability for any two nodes to establish a connection depends on their degrees and can be derived within the \emph{constrained entropy maximization} framework, the score function being represented by Shannon entropy

\begin{equation}\label{equation7}
S=-\sum_{\mathbf{A}}P(\mathbf{A})\ln P(\mathbf{A})
\end{equation}
and the constraints being represented by the degree sequence $\{k_i\}_{i=1}^N$. Upon solving the aforementioned optimization problem \cite{park2004statistical,squartini2011analytical}, one derives the probability that any two nodes establish a connection

\begin{equation}\label{equation7-1}
p_{ij}=\frac{x_ix_j}{1+x_ix_j},\:\forall\:i<j
\end{equation}
the unknowns $\{x_i\}_{i=1}^N$ representing the so-called Lagrange multipliers enforcing the constraints. In order to numerically determine them, one can invoke the \emph{likelihood maximization principle}, prescribing to search for the maximum of the function

\begin{equation}\label{equation8}
\mathcal{L}(\mathbf{x})=\ln P(\mathbf{A}|\mathbf{x})=\ln\left[\prod_{i=1}^N\prod_{j=i+1}^Np_{ij}^{a_{ij}}(1-p_{ij})^{1-a_{ij}}\right]
\end{equation}
with respect to the vector $\{x_i\}_{i=1}^N$, a procedure leading to the resolution of the following system of equations \cite{park2004statistical,squartini2011analytical}

\begin{equation}\label{equation12}
k_i=\sum_{j(\neq i)=1}^Np_{ij}=\sum_{j(\neq i)=1}^N\frac{x_ix_j}{1+x_ix_j},\:\forall\:i.
\end{equation}

\begin{figure}[t!]
\begin{center}
\includegraphics[width=\textwidth]{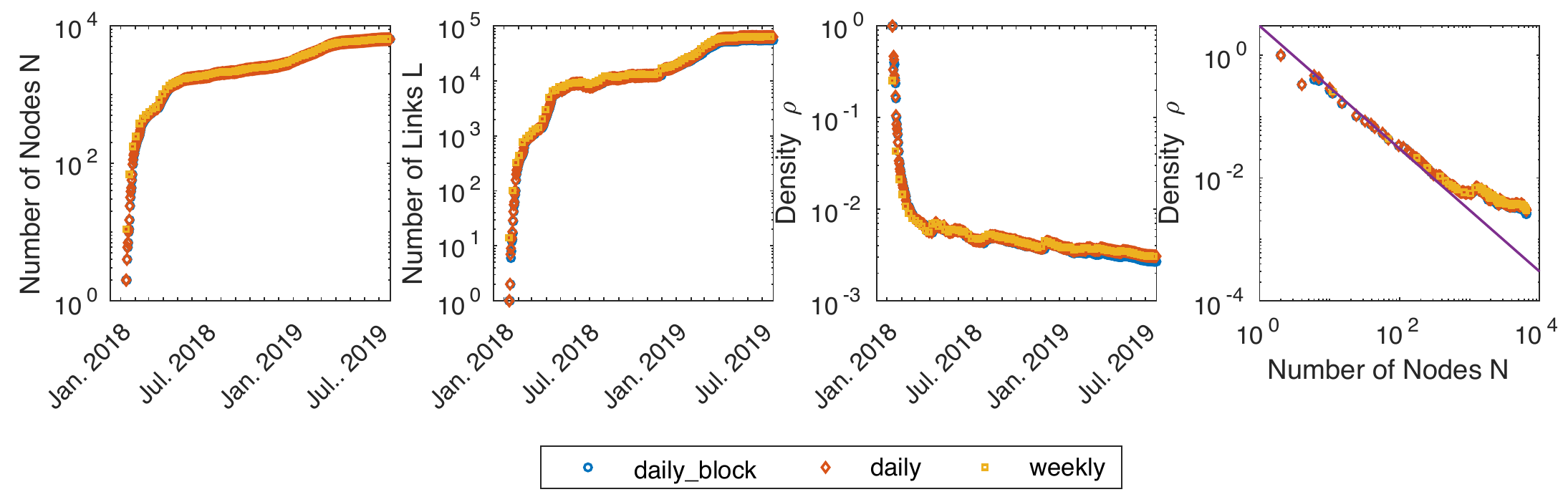}\caption{(colour online) Evolution of the total number of nodes $N$, total number of links $L$ and link density $\rho=\frac{2L}{N(N-1)}$ of the BLN. By plotting the link density versus the total number of nodes, further insight can be gained on the functional dependence $\rho=f(N)$: in particular, the position $\rho\sim N^{-1}$ well describes the link density dependence on $N$ for the snapshots satisfying the condition $N\leq10^3$.}
\label{figure1-1}
\end{center}
\end{figure}

\paragraph{Core-periphery detection.} Inspecting the evolution of centralisation is useful to understand to what extent the structure of a given network becomes increasingly (dis)similar to that of a star graph; however, although encoding the prototypical centralised structure, carrying out a comparison with such a graph can indeed be too simplistic. Hence, we also check for the presence of the ``generalized'' star graph structure also known as \emph{core-periphery structure}, composed by a densely-connected core of nodes surrounded by a periphery of loosely-connected vertices. In order to do so, we implement a recently-proposed approach \cite{dejeude2019detecting}, prescribing to minimize the score function known as \emph{bimodular surprise} and reading

\begin{equation}
S_\parallel=\sum_{i\geq l_c}\sum_{j\geq l_p}\frac{\binom{C}{i}\binom{P}{j}\binom{V-(C+P)}{L-(i+j)}}{\binom{V}{L}}
\end{equation}
where $V=\frac{N(N-1)}{2}$ is the total number of node pairs, $L=\sum_{i=1}^N\sum_{j=i+1}^Na_{ij}$ is the total number of links, $C$ is the number of node pairs in the core portion of the network, $P$ is the number of node pairs in the periphery portion of the network, $l_c$ is the observed number of links in the core and $l_p$ is the observed number of links in the periphery. From a technical point of view, $S_\parallel$ is the p-value of a multivariate hypergeometric distribution \cite{dejeude2019detecting}.

\begin{figure}[t!]
\centering
\includegraphics[width=0.49\textwidth]{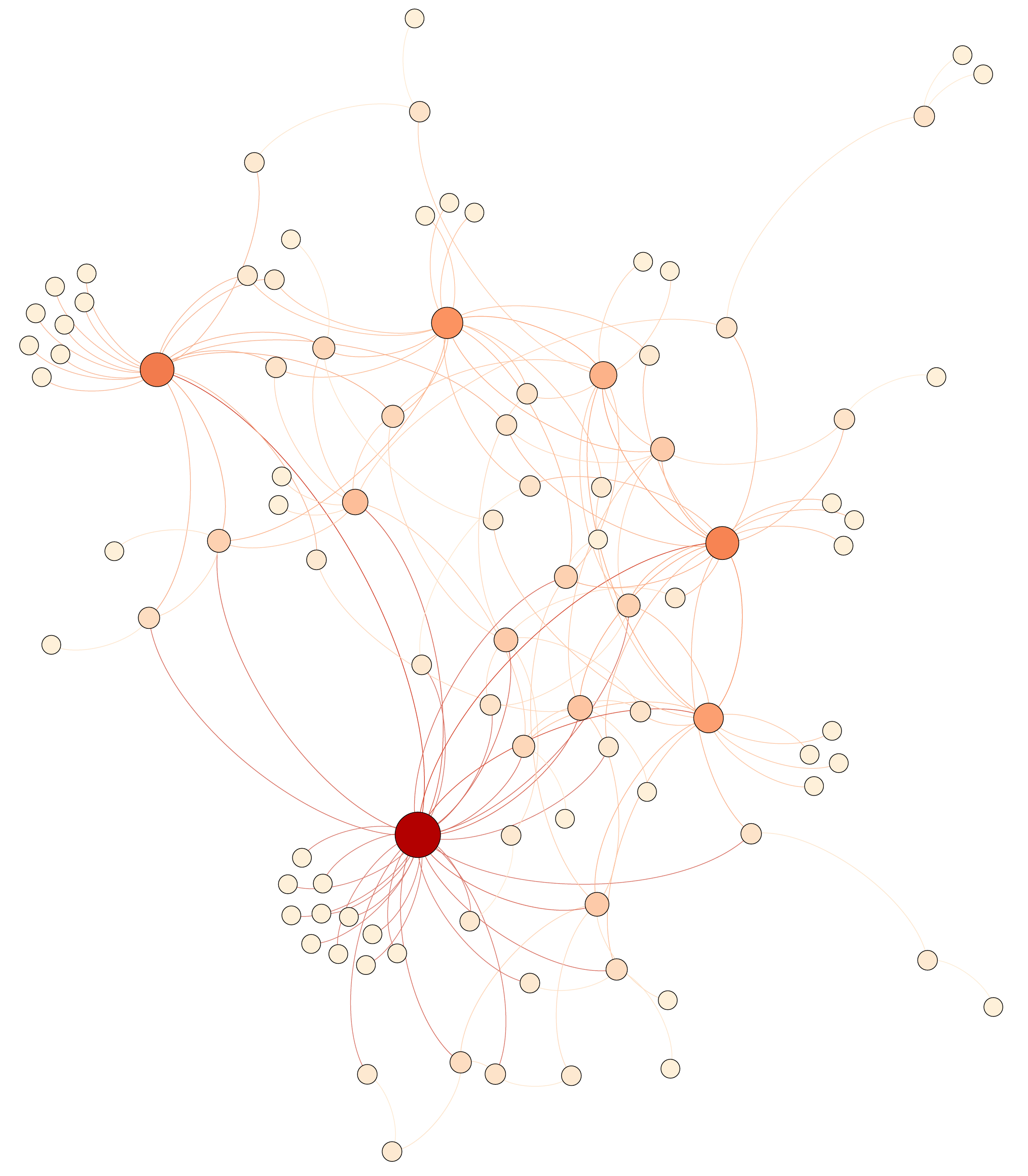}
\includegraphics[width=0.49\textwidth]{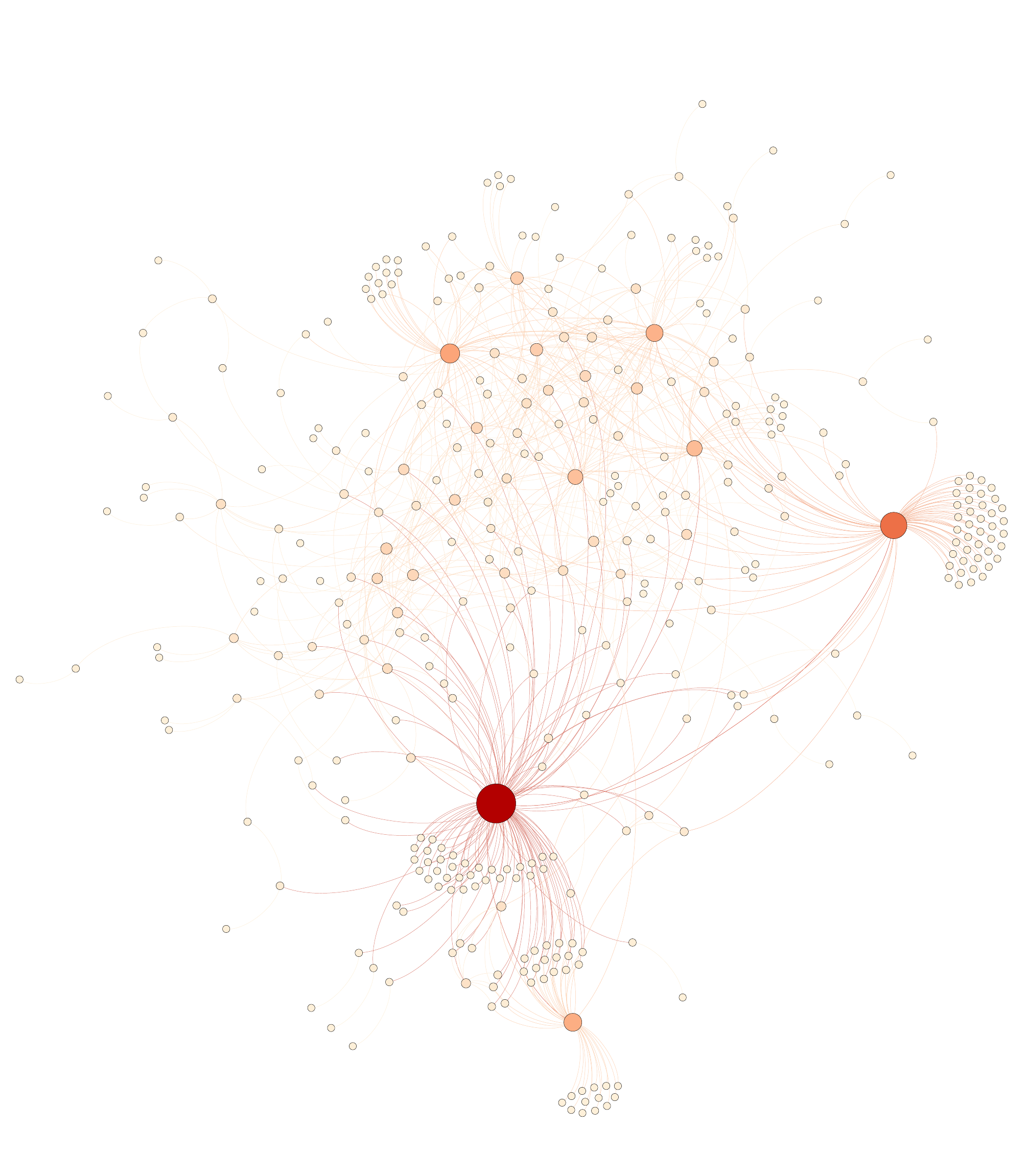}
\caption{(colour online) Comparison between the largest connected component of the BLN (\emph{daily-block snapshot} representation) on day 16 (left - 95 nodes and 155 links are present) and on day 34 (right - 359 nodes and 707 links are present). A visual inspection of the network evolution suggests the presence of a core-periphery structure since its early stages.}
\label{figure1-2}
\end{figure}

\section*{Data}

Since payments in the Bitcoin Lightning Network are \emph{source-routed} and \emph{onion-routed}, the sender must have a reasonably up-to-date view of the network topology, in order to pre-compute the entire payment route. Nodes in the BLN regularly broadcast information about the channels they participate in: each time a channel is opened, or any of its details changes, the two endpoints of the channel announce such changes to the rest of the network. This exchange of information, called \emph{gossip}, allows other nodes to keep their view of the network topology up-to-date, an information that is, then, used to initiate a payment.

The network topology can be visualised by means of the the so-called \emph{routing table}. For this paper, we took regular snapshots of the routing table (every 15 minutes, between January $12^{th}$ 2018, at blockheight 503816, to July $17^{th}$ 2019, at blockheight 585844); these snapshots were, then, aggregated into \emph{timespans}, each timespan representing a constant state of a channel from its start to its end. In addition, this information is enriched with data from the Blockchain: since every channel consists of an unspent transaction output on the Bitcoin Blockchain, we can determine the size of a channel and its open and close dates within minutes. Other heuristics can be used to search for potential channels on the Blockchain, without involving the gossip mechanism: this allows us to put a lower bound on the completeness of our measurements.

In the Bitcoin Blockchain, the time between blocks is Poisson distributed with an expected value of 10 minutes between blocks. On a single day, the expected number of new blocks added to the Blockchain is 144. For the sake of simplicity, and without altering in any way the results, we consider this number of blocks our natural timescale (for example, the blocks of the first day range from the $503816^{th}$ one to the $503959^{th}$ one while the blocks of the second day range from the $503960^{th}$ one to the $504103^{rd}$ one). In this paper, three different representations of the BLN are studied, i.e. the \emph{daily snapshot} one, the \emph{weekly snapshot} one and the \emph{daily-block snapshot} one - even if the results of our analysis will be shown for the daily-block snapshot representation only. A \emph{daily/weekly snapshot} includes all channels that were found to be \emph{active during that day/week}; a \emph{daily-block snapshot} consists of all channels that were found to be \emph{active at the time the first block of the day was released}: hence, the transactions considered for the daily-block representation are a subset of the ones constituting the daily representation.

\begin{figure}[t!]
\begin{center}
\includegraphics[width=\textwidth]{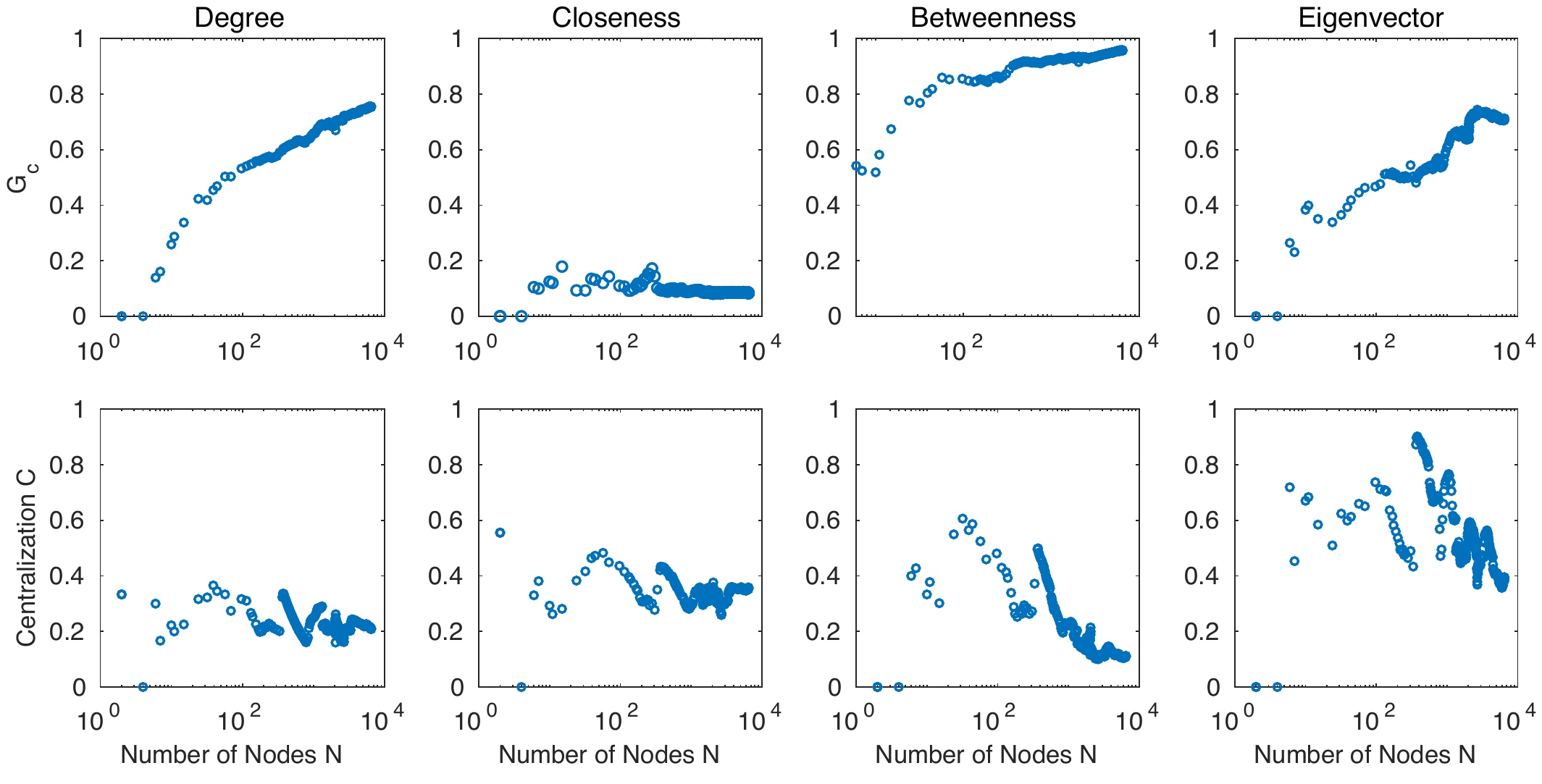}
\caption{(colour online) Top panels: evolution of the Gini index for the degree, closeness, betweenness and eigenvector centrality for the \emph{daily-block snapshot} representation: $G_c$ is characterised by a rising trend, irrespectively from the chosen indicator, pointing out that the values of centrality are increasingly unevenly distributed. Bottom panels: evolution of the degree-, closeness-, betweenness- and eigenvector-centralisation measures: although the eigenvector-centralization index reaches quite large values in the middle stages of the BLN history, the picture provided by a star graph is too simple to faithfully represent the BLN structure.}
\label{figure1-3}
\end{center}
\end{figure}

\section*{Results}

\paragraph{Empirical analysis of the BLN binary structure.} Figure \ref{figure1-1} plots the evolution of basic network quantities since launch of the BLN, i.e. the number of nodes, which is a proxy of the number of users, the number of links and the link density. As it can be seen, although the network size increases (for the \emph{daily-block snapshot} $N$ ranges from $2$ to $6476$ and $L$ ranges from $1$ to $55866$; in particular, in the last daily snapshot of our dataset we have 6476 nodes and 54440 links), it becomes sparser. However, two different regimes are visible: a first phase where a steep increase of $N$ and $L$ (descrease of $\rho$) takes place is followed by a phase during which a much smoother increase (decrease) of the same quantities is observed. Further insight on the BLN evolution can be gained by plotting the link density $\rho=\frac{2L}{N(N-1)}$ versus the total number of nodes $N$: a trend whose functional form reads $\rho\sim cN^{-\gamma}$, with $\gamma\simeq 1$, clearly appears. However, such a functional form seems to describe quite satisfactorily the BLN evolution up to the period when $N\simeq10^3$: afterwards, a different functional dependence seems to hold. Notice also that the value of the numerical constant $c$ coincides with the value of the average degree, since $c=\frac{2L}{N-1}=\frac{\sum_{i=1}^Nk_i}{N-1}\simeq\overline{k}$. By imagining a growth process according to which each new node enters the network by establishing at least one new connection with the existing ones, to ensure that $L^t\geq N^t-1$, a lower-bound on $c\simeq\overline{k}$ can be deduced: $c\geq2$ (fig. \ref{figure1-1} shows the trend $y=3N^{-1}$ even if the inspection of the evolution of the quantity $c=\frac{2L}{N-1}$ reveals that periods where $c\simeq\overline{k}$ assumes different, constant values can be individuated).

\begin{figure}[t!]
\begin{center}
\includegraphics[width=\textwidth]{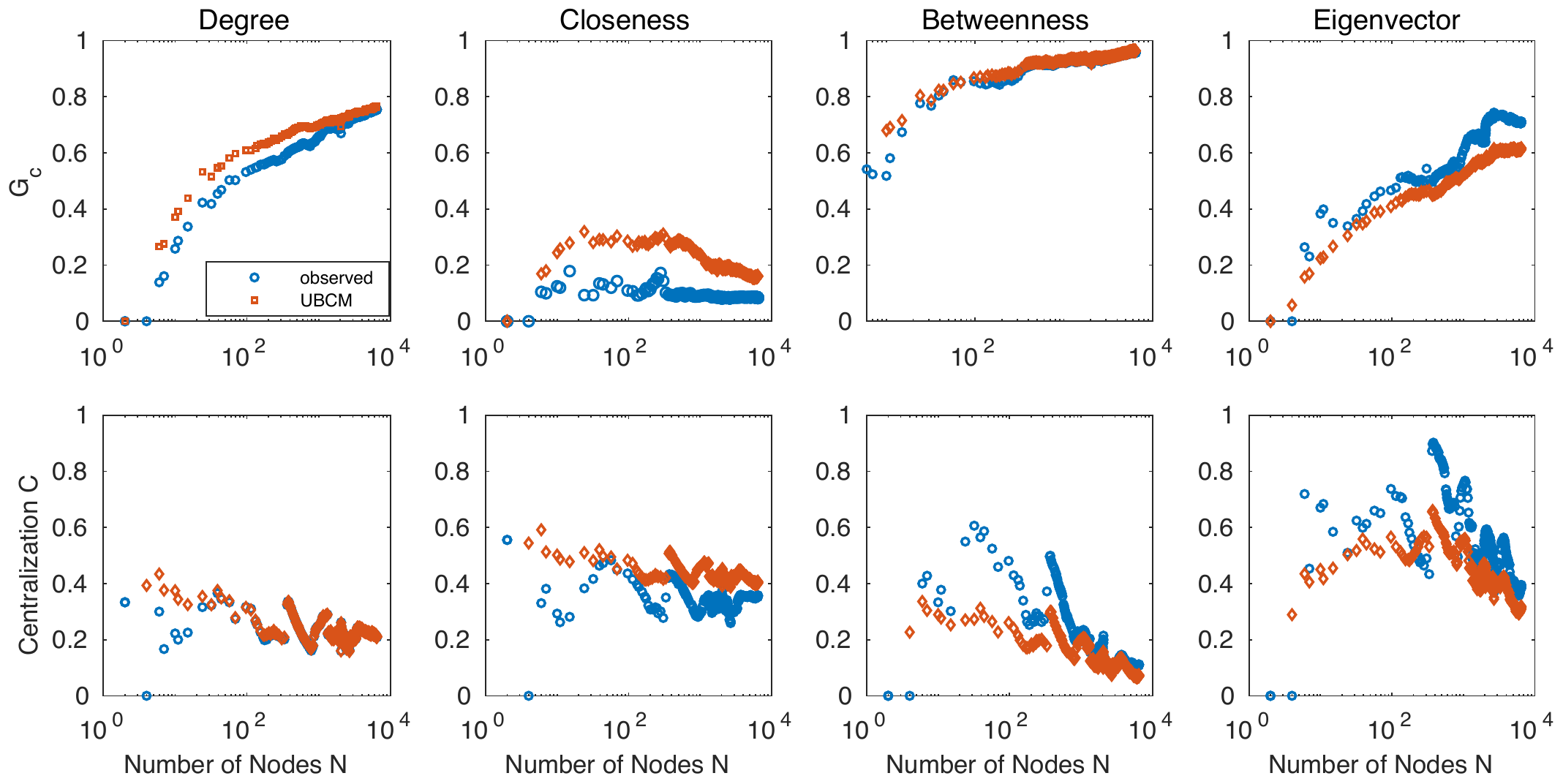}
\caption{(colour online) Top panels: comparison between the observed Gini index for the degree, closeness, betweenness and eigenvector centrality (blue dots) and their expected value, computed under the UBCM (red diamonds) for the \emph{daily-block snapshot} representation. Once the information contained into the degree sequence is properly accounted for, a (residual) tendency to centralisation is still visible. Bottom panels: comparison between the observed degree-, closeness-, betweenness- and eigenvector-centralisation measures and their expected value computed under the UBCM (red diamonds). Once the information contained into the degree sequence is properly accounted for, the emerging picture is that of a network characterized by some kind of more-than-expected star-likeness: deviations from this benchmark, however, are clearly visible and probably due to the co-existence of many star-like sub-structures (see also fig. \ref{figure1-2}).}
\label{figure1-4}
\end{center}
\end{figure}

In order to comment on the centrality structure of the BLN, let us explicitly draw it: fig. \ref{figure1-2} shows the largest connected component of the BLN daily-block snapshot representation on day 16 and on day 34. Several hubs are present (e.g. on day 34, the largest one, having degree $k_{hub}^{34}=121$, is linked to the $34.3\%$ of nodes): notice that each of them is linked to a plethora of other nodes that, instead, are scarcely linked among themselves. The emergence of structurally-important nodes is further confirmed by plotting the evolution of the Gini index for the distribution of the centrality measures defined in the Methods section (i.e. the degree, the closeness, the betweenness and the eigenvector centrality): fig. \ref{figure1-3} shows that $G_c$ is increasing for three measures out of four, pointing out that the values of centrality are more and more unevenly distributed (irrespectively from the chosen indicator). The flat trend characterizing the closeness centrality could be explained by the presence of nodes with large degree ensuring the vast majority of nodes to be reachable quite easily. On the other hand, the evolution of the centralisation indices indicates that the BLN is \emph{not} evolving towards a star graph, although the eigenvector centrality reaches quite large values in the middle stages of the BLN history. As anticipated above, imagining that the picture provided by a star-like structure could provide a good description of the BLN topology is indeed too simplistic.

\paragraph{Benchmarking the observations.} Let us now benchmark the observations concerning the centrality and the centralisation indices with the predictions for the same quantities output by the UBCM. More specifically, we have computed the expected value of $G_c$ and $C_c$ (with $c_i=k^c_i,c^c_i,b^c_i,e^c_i,\:\forall\:i$) and the corresponding error, by explicitly sampling the ensembles of networks induced by the UBCM. In fig. \ref{figure1-4} we plot and compare the evolution of the observed and expected values of $G_c$ and $C_c$, both as functions of $N$. Such a comparison reveals that the UBCM tends to overestimate the values of the Gini index for the degree, the closeness and the betweenness centrality and to underestimate the values of the Gini index for the eigenvector centrality\footnote{Z-scores, not shown here, confirm that all observations are statistically significant.}. These results point out a behavior that is not reproducible by just enforcing the degree sequence (irrespectively from the chosen index). The evidence that the UBCM predicts a more-heterogeneous-than observed structure, could be explained starting from the result concerning the eigenvector centrality. The latter, in fact, seems to indicate a non-trivial (i.e. not reproducible by lower-order constraints like the degrees) tendency of well-connected nodes to establish connections among themselves - likely, with nodes having a smaller degree attached to them. Such a \emph{disassortative} structure could explain the less-than-expected level of unevenness characterizing the other centrality measures: in fact, each of the nodes behaving as the ``leaves'' of the hubs would basically have the same values of degree, closeness and betweenness centrality.

On the other hand, the betweenness- and the eigenvector-centralisation indices suggest that the BLN structure is indeed characterized by some kind of more-than-expected star-likeness: the deviations from the picture provided by such a benchmark, however, could be explained by the co-existence of \emph{multiple} star-like sub-structures (see also fig. \ref{figure1-2} and the Appendix for a more detailed discussion about this point).

\begin{figure}[t!]
\centering
\includegraphics[width=0.49\textwidth]{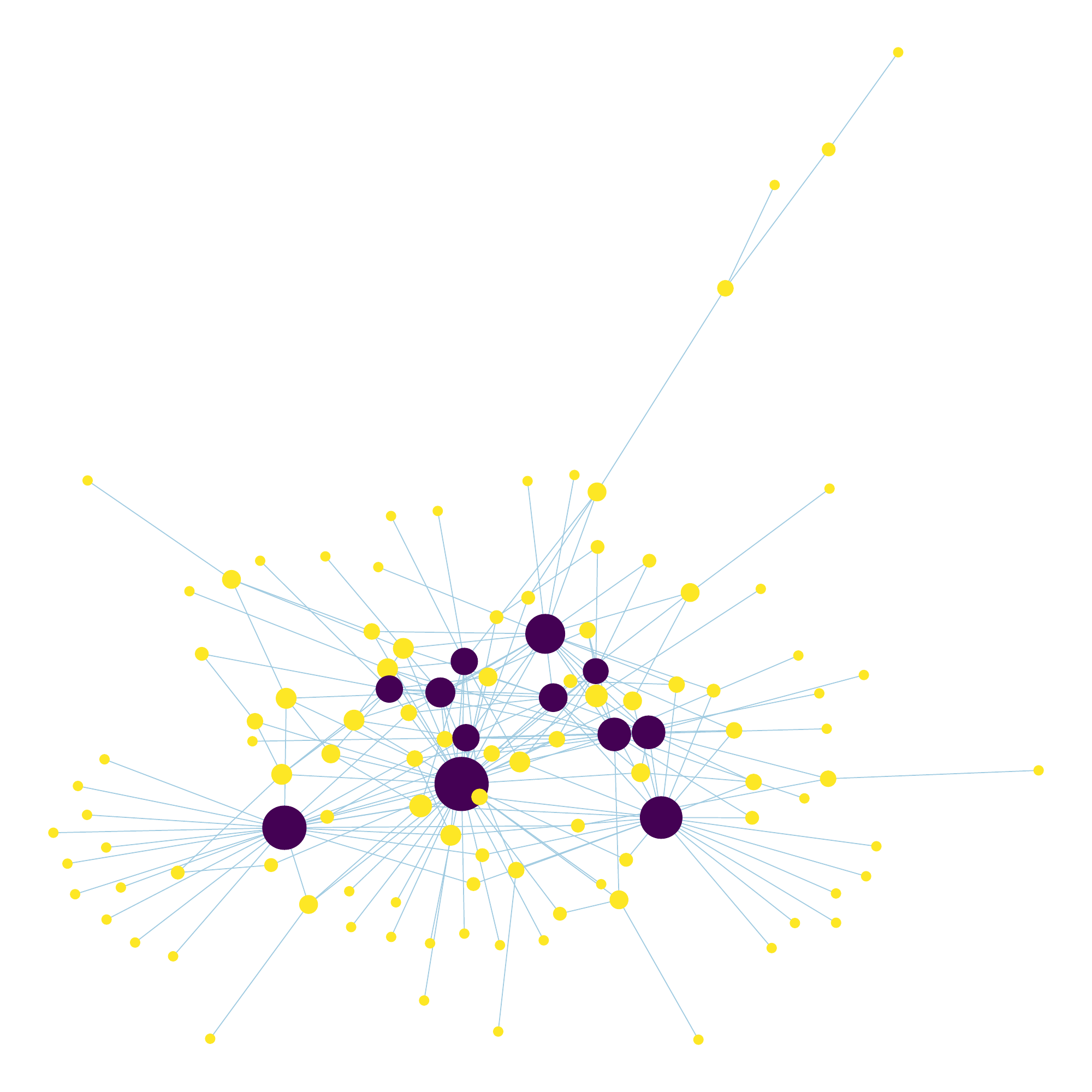}
\includegraphics[width=0.49\textwidth]{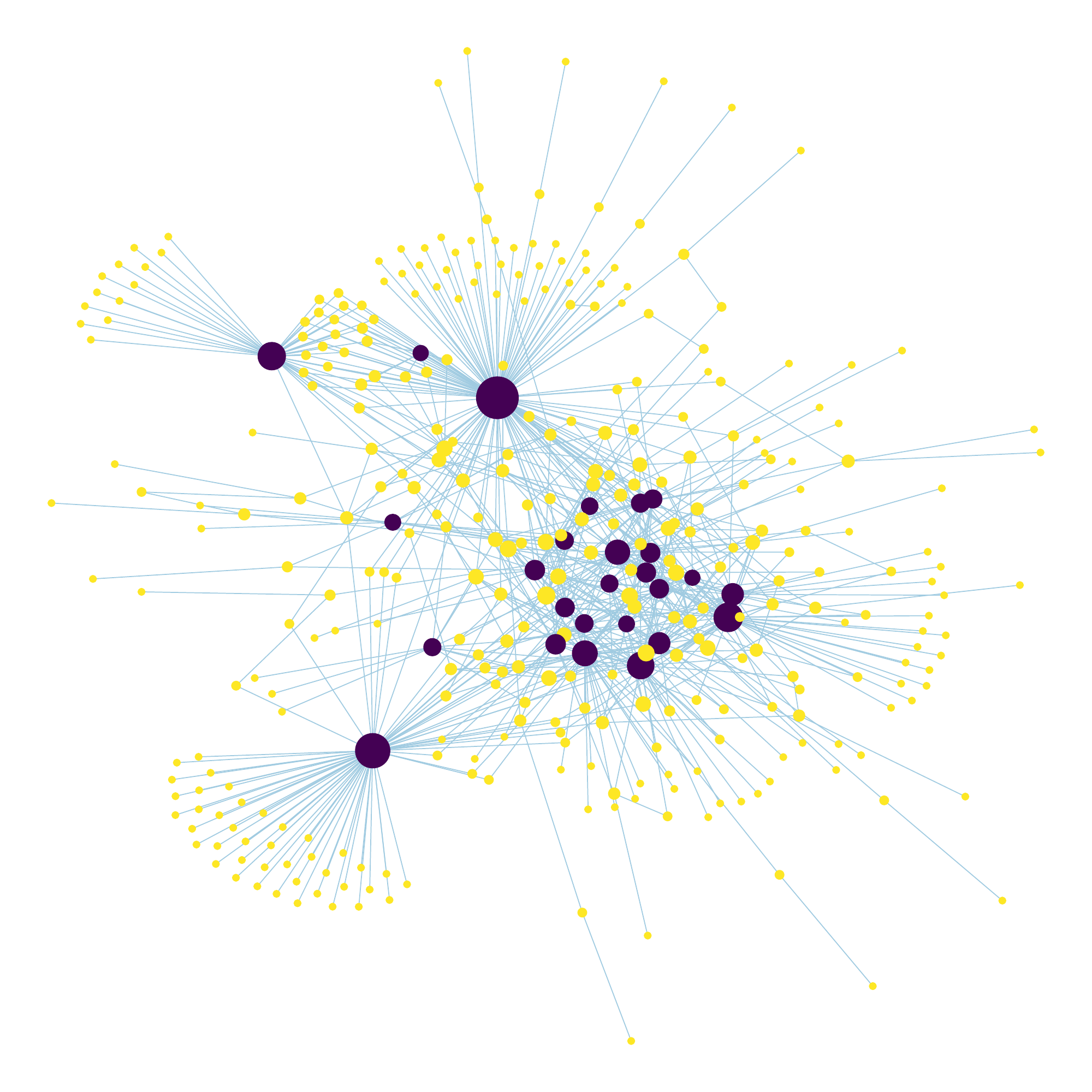}
\caption{(colour online) Core-periphery structure of the BLN \emph{daily-block snapshot} representation on day 16 (left - 95 nodes and 155 links are present) and on day 34 (right - 359 nodes and 707 links are present), with core-nodes drawn in blue and periphery-nodes drawn in yellow.}
\label{figure1-5}
\end{figure}

\paragraph{Core-periphery detection.} A clearer picture of the BLN topological structure is provided by the analysis aimed at clarifying the presence of a ``core-periphery -like'' organization. Inspecting the evolution of the bimodular surprise $S_\parallel$ across the entire considered period reveals that the statistical significance of the recovered core-periphery structure increases, a result leading to the conclusion that the description of the BLN structure provided by such a model becomes more and more accurate as the network evolves. As an example, fig. \ref{figure1-5} shows the detected core-periphery structure on the snapshots depicted in fig. \ref{figure1-2}: the nodes identified as belonging to the core and to the periphery are, respectively, coloured in blue and yellow. 

\begin{figure}[t!]
\begin{center}
\includegraphics[width=\textwidth]{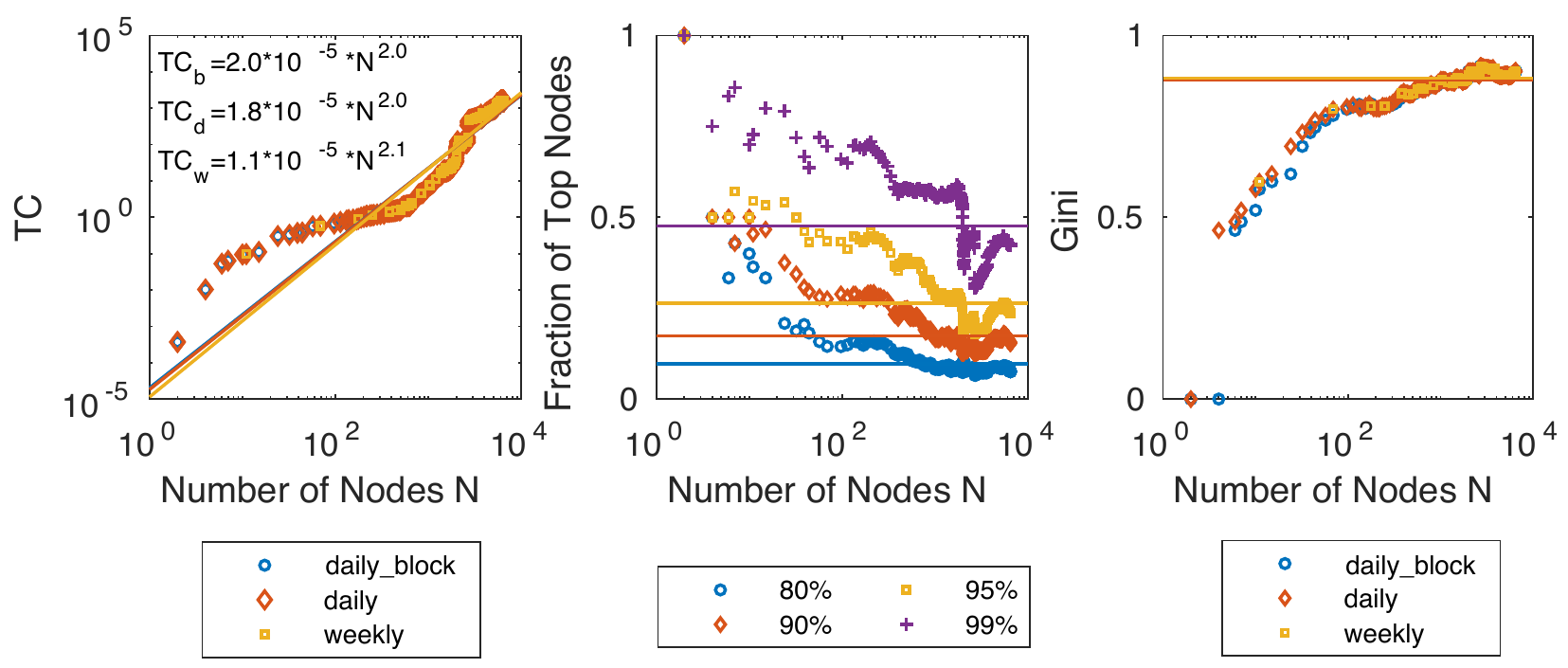}
\caption{(colour online) Evolution of the total capacity of the BLN (left). Percentage of nodes holding the $\simeq80\%$, $\simeq90\%$, $\simeq95\%$ and $\simeq99\%$ of the total number of bitcoins at stake in the BLN (middle): the former has been computed as the fraction $\frac{n^*}{N}$ of top nodes whose total capacity amounts at $\simeq80\%$, $\simeq90\%$, $\simeq95\%$, $\simeq99\%$ of the total. Evolution of the Gini coefficient $G_s$ (right): although the total number of bitcoins rises, inequality rises as well.}
\label{figure1-6}
\end{center}
\end{figure}

\paragraph{Empirical analysis of the BLN weighted structure.} Let us now move to the empirical analysis of the weighted structure of the BLN, by inspecting the evolution of the total capacity $W$ of (i.e. the total number of bitcoins within) the BLN daily-block snapshot representation: fig. \ref{figure1-6} shows the evolution of $W$ as a function of network size $N$. The trend shown in the same figure reads $y=aN^b$ with $a=2\cdot10^{-5}$ and $b=2$. Although the total number of bitcoin rises, inequality rises as well: in fact, the percentage of nodes holding a given percentage of bitcoins at stake in the BLN steadily decreases (on average, across the entire period, about the $10\%$ ($50\%$) of the nodes holds the $80\%$ ($99\%$) of the bitcoins - see the second panel of fig. \ref{figure1-5}). This trend is further confirmed by the evolution the Gini coefficient $G_s$, whose value is $\simeq0.9$ for the last snapshots of our dataset (and whose average value is $0.88$ for the daily-block snapshot representation).

\section*{Conclusions}

The Bitcoin Lightning Network is a sort of ``Layer 2'' protocol aimed at speeding up the Blockchain, by enabling fast transactions between nodes. Originally designed to allow for cheaper and faster transactions without sacrificing the key feature of Bitcoin, i.e. its decentralisation, it is evolving towards an increasingly centralised architecture, as our analysis reveals. In particular, many star-like sub-structures, whose centers coincide with the ``centrality hubs'' revealed by the Gini coefficient, co-exist. These hubs act as channel-switching nodes and seem to emerge as an unavoidable consequence of the way BLN is designed: as a route through the network must be found and longer routes are more expensive (fees are present for the gateway service provided by intermediate nodes), any two BLN users will search for a short(est) path. At the same time, nodes (which can only create channels based on local information) have the incentive to become as central as possible within the BLN, in order to maximize the transaction fees they may earn. Hubs may, thus, have emerged as a consequence of the collective action of users following the two aforementioned behaviors - and, from this perspective, it is not surprising that central nodes have been observed since the very beginning of the BLN history.

For what concerns hubs interconnectedness, then, previous results have shown that mechanisms of centrality-maximizing agents yield a core-periphery structure \cite{Konig2010,Konig2014} (regardless of the notion of centrality the agents attempt to maximize), an evidence indicating that the presence of both topological signatures can be compactly inspected by studying (the evolution) of eigenvector centrality. As a last observation, we also notice that the presence of ``centrality hubs'' seems to be at the origin of another structural BLN peculiarity, i.e. its small-world -ness (a feature already revealed by previous studies \cite{rohrer2019discharged}).

The tendency to centralisation is observable even when considering weighted quantities, as only about $10\%$ ($50\%$) of the nodes hold $80\%$ ($99\%$) of the bitcoins at stake in the BLN (on average, across the entire period); moreover, the average Gini coefficient of the nodes strengths is $\simeq0.88$. These results seems to confirm the tendency for the BLN architecture to become ``less distributed'', a process having the undesirable consequence of making the BLN increasingly fragile towards attacks and failures.

\section*{Authors contributions}

Jian-Hong Lin and Kevin Primicerio performed the analysis. Tiziano Squartini, Christian Decker and Claudio J. Tessone designed the research. All authors wrote, reviewed and approved the manuscript.

\section*{Acknowledgements}

Jian-Hong Lin acknowledges support from the China Scholarship Council (no. 2017083010177).

\section*{Appendix}

\begin{figure}[t!]
\begin{center}
\includegraphics[width=0.39\textwidth]{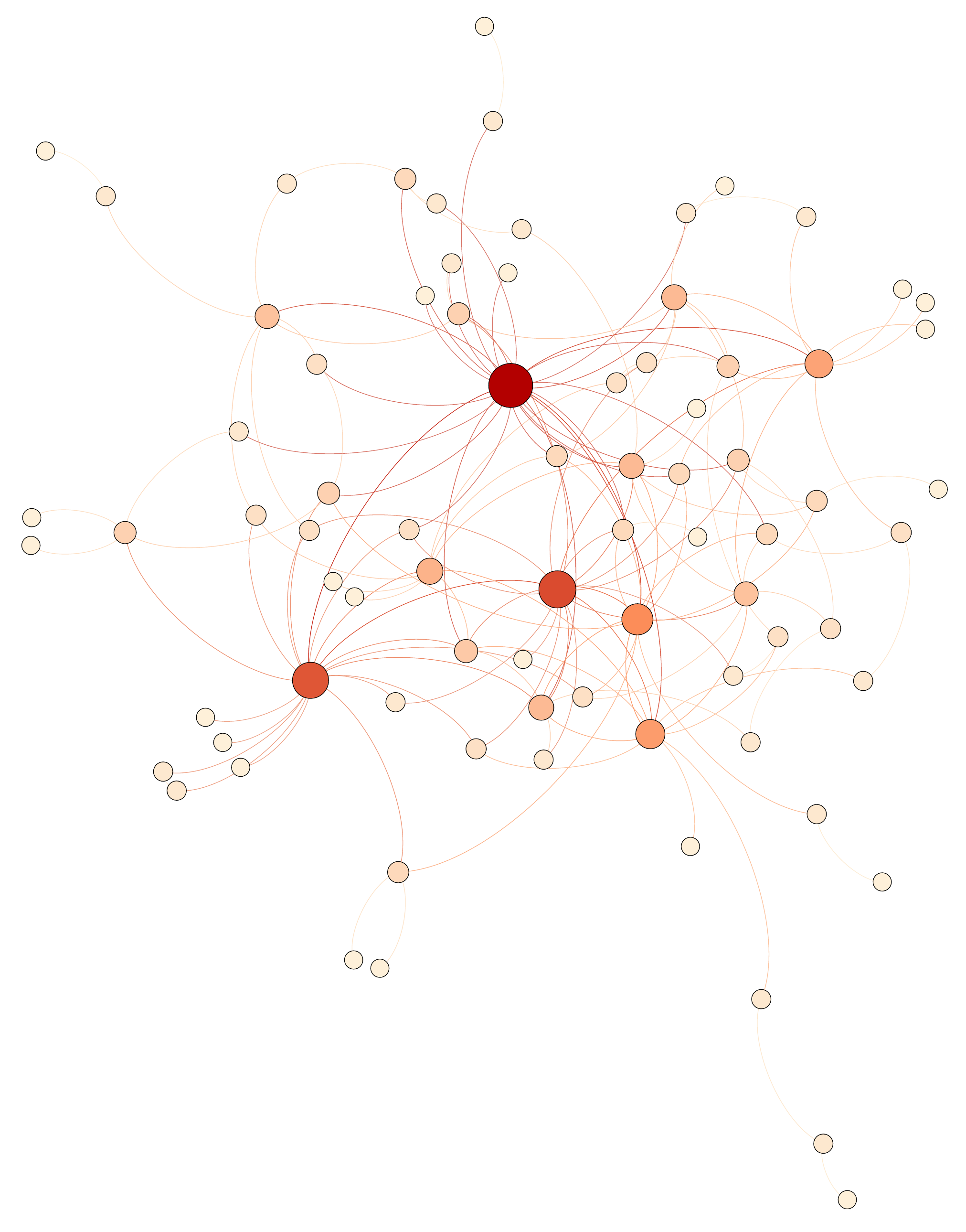}
\includegraphics[width=0.59\textwidth]{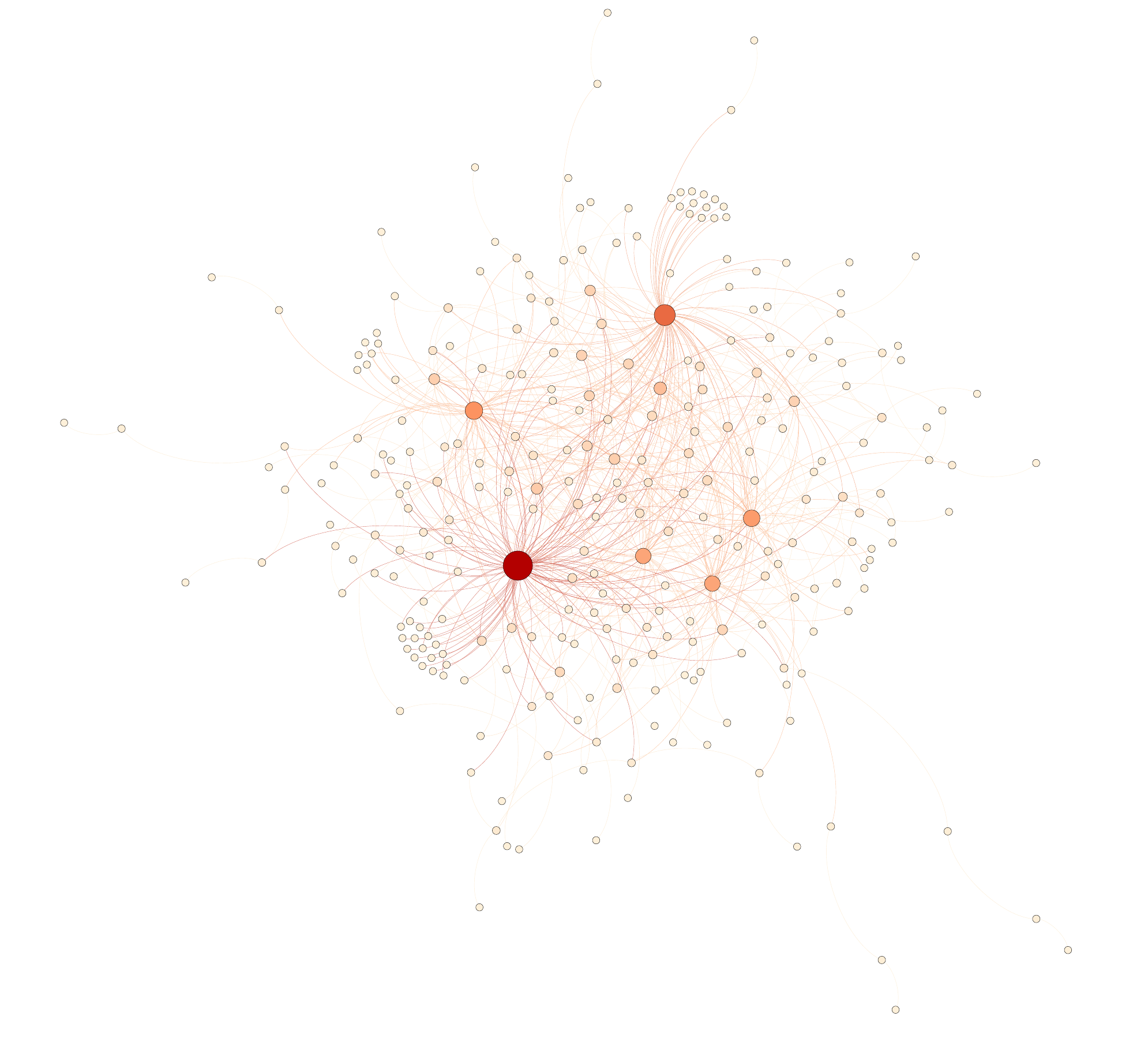}
\includegraphics[width=0.49\textwidth]{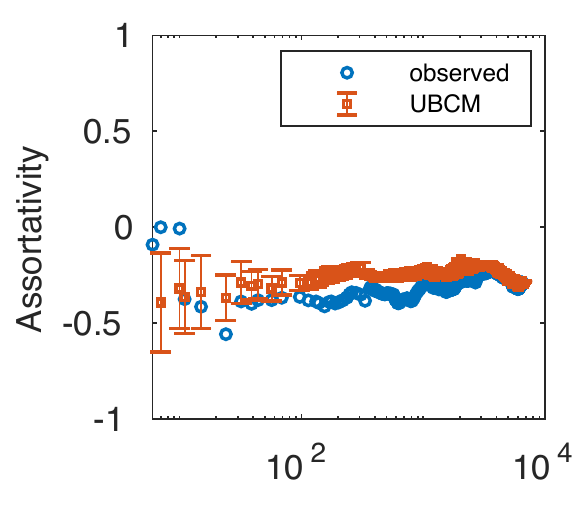}
\caption{(colour online) Top panels: comparison between the largest connected component of the BLN (\emph{daily-block snapshot} representation) generated by the UBCM for the day 16 and the day 34. A visual inspection of these networks confirms that star-like sub-structures are present to a much lesser extent with respect to the observed BLN in the same snapshots. Bottom panel: evolution of the comparison between the empirical assortativity coefficient $r$ (blue dots) and its expected value, computed under the UBCM (red diamonds), for the \emph{daily-block snapshot} representation. The BLN is significantly more disassortative than expected.
}
\label{figure1-7}
\end{center}
\end{figure}

As anticipated in the main text, the UBCM seems to underestimate the extent to which the topological structure of the BLN is disassortative. Figure \ref{figure1-7} shows the evolution of the Newman \emph{assortativity coefficient} \cite{newman2002assortative}, defined as

\begin{equation}
r=\frac{L\sum_{i=1}^N\sum_{j(\neq i)=1}^Na_{ij}k_ik_j-\left(\sum_{i=1}^Nk_i^2\right)^2}{L\sum_{i=1}^Nk_i^3-\left(\sum_{i=1}^Nk_i^2\right)^2};
\end{equation}
and its expected counterpart under the UBCM: as it is clearly visible, the BLN is more disassortative than expected (i.e. the correlations between degrees are ``more negative'' than predicted by the UBCM), the reason lying in the presence of the aforementioned star-like sub-structures that, instead, are absent in the model. To further confirm this, we explicitly show two configurations drawn from the UBCM for the snapshots 16 and 34: as fig. \ref{figure1-7} clearly shows, star-like sub-structures are present to a much lesser extent with respect to the observed counterparts shown in fig. \ref{figure1-2}.

\bibliography{LN}
\end{document}